\documentclass[titlepage,twocolumn]{revtex4-1}

\usepackage{bbold}
\usepackage{amssymb}
\usepackage{graphics}
\usepackage{graphicx}
\usepackage{epstopdf}
\usepackage{multirow}
\usepackage{amsmath}
\usepackage{amsthm}
\usepackage{amstext}
\usepackage{amscd}
\usepackage{epsfig}
\usepackage[mathscr]{eucal}
\usepackage{times}
\usepackage{srcltx}
\usepackage{shadow}
\usepackage{url}
\usepackage{color}
\usepackage[all]{xy}

\usepackage[most]{tcolorbox}

\usepackage{hyperref}

\begin{document}

\title{Evolution of chemotactic hitchhiking}


\author{Gurdip Uppal${}^{1}$, Weiyi Hu${}^{1,2}$, Dervis Can Vural${}^{1}$}
\email[]{dvural@nd.edu}
\affiliation{${}^{1}$Department of Physics, University of Notre Dame, USA \\
${}^{2}$Department of Mathematics, Sichuan University, China }


\date{\today}
\begin{abstract}
Bacteria typically reside in heterogeneous environments with various chemogradients where motile cells can gain an advantage over non-motile cells. Since motility is energetically costly, cells must optimize their swimming speed and behavior to maximize their fitness. Here we investigate how cheating strategies might evolve where slow or non-motile microbes exploit faster ones by sticking together and hitching a ride. Starting with physical and biological first-principles we computationally study the effects of sticking on the evolution of motility in a controlled chemostat environment. We find stickiness allows slow cheaters to dominate when nutrients are dispersed at intermediate distances. Here, slow microbes exploit faster ones until they consume the population, leading to a tragedy of commons. For long races, slow microbes do gain an initial advantage from sticking, but eventually fall behind. Here, fast microbes are more likely to stick to other fast microbes, and cooperate to increase their own population. We therefore find the nature of the hitchhiking interaction, parasitic or mutualistic, depends on the nutrient distribution. 
\end{abstract}

\maketitle


\section*{Introduction}

Microbial motility plays an essential role in biofilm formation \cite{guttenplan2013regulation}, dispersal \cite{kaplan2010biofilm}, virulence \cite{josenhans2002role}, and biogeochemical processes \cite{stocker2012ecology}. However, motility also comes with metabolic \cite{mitchell2002energetics, yi2016phenotypic} and ecological \cite{stocker2012ecology} costs and can also increase the rate of predation \cite{gerritsen1977encounter} and viral infection \cite{Murray1992Viral}. Between such costs and benefits, evolution optimizes how to, when to, where to, and how fast to swim.





Interestingly, motility also promotes aggregation and adhesion of microbes \cite{alexandre2015chemotaxis}. There appears to be a connection between the metabolic pathways that regulate chemotaxis and those that regulate clumping behavior \cite{bible2008function, piepenbrink2016motility}. 
Furthermore, non-motile bacteria can stick to motile ones, thereby dispersing without paying the energetic cost. This phenomenon is known as microbial hitchhiking. Hitchhiking behavior has been observed in a variety of microbial species \cite{samad2017swimming, Miller2019metabolic, shrivastava2018cargo}, where the co-dynamics of motile and non-motile species can lead to pattern formation \cite{xiong2019flower}. In the oral microbiome, multiple non-motile species of bacteria hitchhike gliding bacteria \emph{C. gingivalis} to disperse and shape the spatial diversity of the microbiome \cite{Miller2019metabolic, shrivastava2018cargo}. In ocean environments, zooplankton can transport microbes across otherwise untraversable strata \cite{grossart2010bacteria}. Co-swarming of motile species can also allow species to combine their skills \cite{venturi2010co}.

However, the ecology of microbial hitchhiking is not entirely clear. For example, some experimental studies of the swarming bacteria \emph{P. vortex} aiding the dispersal of non-motile microbes seem to suggest a possible mutualistic relationship. In experiments with the non-motile \emph{X. perforans}, \emph{X. perforans} attracted and directed the motility of \emph{P. vortex} to facilitate its own dispersal \cite{hagai2014surface}. In another case, \emph{P. vortex} helped transport conidia of the filamentous fungus \emph{A. fumigatus} \cite{ingham2011mutually} and were able to rescue fungal spores from areas harmful to the fungus. Here, the bacteria also gain an advantage by utilizing germinating mycelia as bridges to cross air gaps \cite{warmink2011hitchhikers, warmink2009migratory}. In another experiment, \emph{P. vortex} carried non-motile \emph{E. coli} strain as cargo to help degrade antibiotics \cite{finkelshtein2015bacterial}. Here \emph{P. vortex} used a bet hedging strategy where it only carried the cargo bacteria when needed. In this case, the motile \emph{P. vortex} actually seemed to gain more from the relationship. Both species had an increase in population when co-inoculated but \emph{P. vortex} grew $10^8$-fold compared to \emph{E. coli} which increased 100-fold \cite{finkelshtein2015bacterial}.

Though hitchhiking behavior has been observed in a variety of microbial species, a theoretical understanding incorporating the effects of cell and nutrition density, propensity to stick, and the hydrodynamic interactions between microbes is lacking. 

Here we fill this gap by studying the evolution of swimming strategies of chemotactic microbes that interact with each other and the habitat fluid through contact and hydrodynamic forces.  Studies have shown hydrodynamic effects depending on microbial shape, swimming mechanism, and interactions with boundaries can strongly influence swimming patterns \cite{lauga2009hydrodynamics}. Hydrodynamic interactions between microbes have been shown to promote aggregation in spherical \cite{ishikawa2008coherent} and rod shaped bacteria \cite{saintillan2007orientational}. Sperm cells can aggregate to better align and increase their overall velocity \cite{fisher2014dynamics}. The shape of cells will also determine the convective \cite{zhan2014accumulation} and drag \cite{lee2019hydrodynamic,filippov2000drag} forces. It has also been shown that pair-wise swimming is not stable without extra aggregation mechanisms \cite{ishikawa2007hydrodynamic}. Interactions with self-generated flows can also drastically effect motility. Fluid flows created by bacteria can drive self organization \cite{lushi2014fluid} and influence chemotactic motion \cite{lushi2012collective}. Despite these developments, the evolutionary and ecological consequences of hydrodynamic and contact forces between motile microorganisms have not been explored. 

Our goal is to start with the physics of flow, drag and aggregation, and from here draw ecological and evolutionary implications on the emergence of social and anti-social behavioral strategies in microbial swarms. Specifically, in our evolutionary simulations, we account for the possibility that two microbes might temporarily stick upon colliding; for their indirect pushes, pulls, and torques on each other from a distance (as their swimming alters the fluid flow surrounding them); and for the difference in frictional  (drag) forces exerted by the fluid when they are swimming in solitude, versus stuck together.



One common mode of motility in the bacterial world is run-and-tumble chemotaxis, where bacteria  perform a random walk with step lengths that depend on the local concentration gradient along the swim direction \cite{celani2010bacterial, chatterjee2011chemotaxis}. The bacteria will run straight for a longer duration if the nutrients or toxins are changing favorably; if not, it will keep the run short and ``tumble'', picking a new random direction. In the present work too, we model the evolution of microbes carrying out this mode of chemotaxis. 
 
In summary, our model assumptions, stated qualitatively, are as follows: (1) Microbes perform  run-and-tumble chemotaxis \cite{celani2010bacterial, chatterjee2011chemotaxis}, for which we use precise chemotaxis response functions derived from empiric data \cite{segall1986temporal, berg1972chemotaxis} (2) Microbes are  placed in the low end of a nutrition gradient every $\tau$ hours, as if in an evolution experiment, or as if in a still fluid body in which resources appear at a certain distance every $\tau$ hours. We call each such time interval, a ``race''. (3) Microbes pay a metabolic cost directly proportional to their run speed, and run speed is heritable (4) the microbial growth rate at a given location depends on the nutrient concentration at that location, in accordance with empiric data \cite{monod1949growth, gibson2018distribution} (5) when two microbes collide, they either stick, or not stick depending on the stickiness trait of the microbes. We study three separate cases: (a) all microbes are sticky, (b) no microbe is sticky, (c) only some microbes are sticky, and stickiness is heritable except for random mutations. (6) swimming microbes alter the flows surrounding them, which causes them to push, pull, and reorient each other. The microbes also experience drag (friction) force from the fluid, which is different for stuck and unstuck microbes. The spatial and angular form of physical forces used in our simulations are gathered from first-principles experimental and computational fluid dynamics studies \cite{lauga2009hydrodynamics, berke2008hydrodynamic, lee2019hydrodynamic, filippov2000drag}.

Operating under these assumptions, we find that  (1) For short races, the best strategy for everyone is to not swim. (2) For intermediate-length races, sticking allows slow runners to make it to large nutrition concentrations without exerting much effort themselves and exploiting the motility of faster swimmers. (3) For long races, fast runners are ultimately able to leave behind the slow hitchhikers regardless of whether they themselves are sticky, and ultimately dominate the population. In summary, we find that stickiness selects for hitchhiking behavior only for races that are intermediately long.

We also find that (4) the evolutionarily stable strategy will sensitively depend on the initial distribution of run speed and stickiness as well as the difference between drag forces experienced by stuck versus unstuck microbes. We find that when the drag experienced by a stuck pair is sufficiently lower than the drag on two individuals (which is a matter of microbial shape), fast microbes develop cooperation by sticking together and further increasing their net speed. In this case, sticking goes from mediating a parasitic interaction (leading to tragedy of commons, i.e. all microbes slowing down) for intermediate-length races, to an evolutionarily stable mutualistic interaction amongst fast microbes for long races.


This paper is outlined as follows, we first study the effects of sticking when all microbes are equally sticky, and stickiness does not mutate. We then study the effects of varying microbial density and steepness of the chemogradient. Finally, we study how sticking strategies evolve depending on microbe shape (i.e. depending on how drag forces change as microbes pair up) as well as when sticking is a evolvable trait.



\section*{Methods}
\begin{figure}
    \centering
    \includegraphics[width=0.49\textwidth]{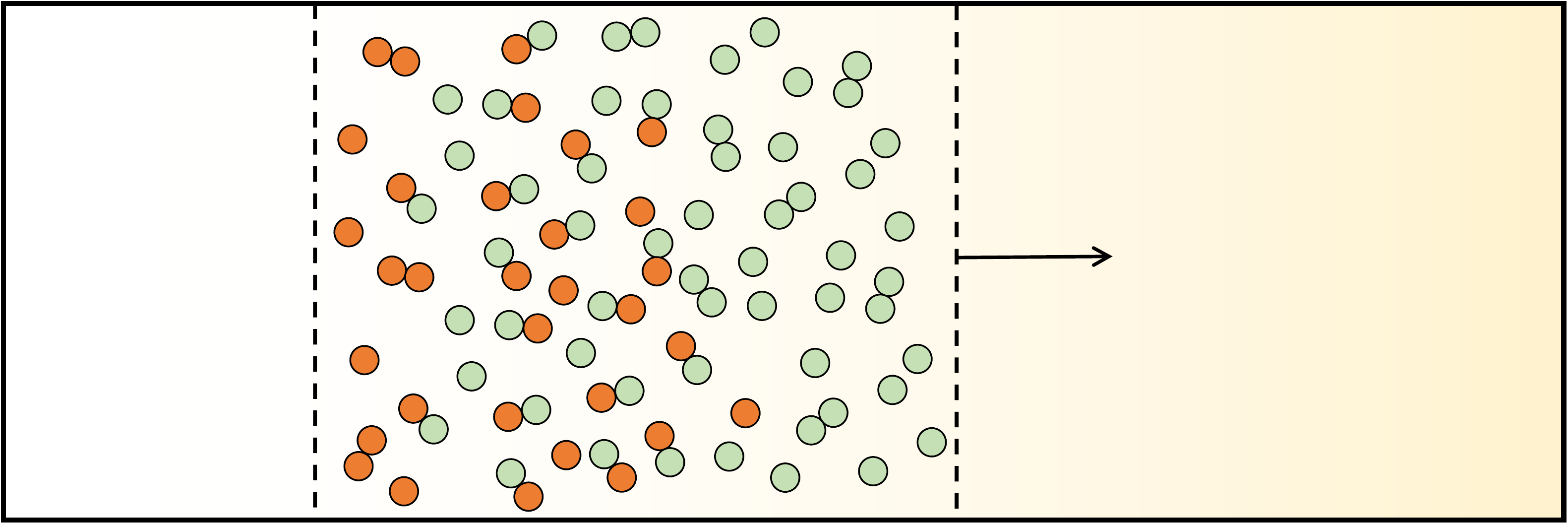}
    \caption{{\bf Model schematics.} A band of microbes perform run-and-tumble chemotaxis in a channel with linear chemoattractant gradient (yellow). Colliding microbes stick together until the next tumble. Sticking can be beneficial, since pairs experience less drag force. However on the flip side, slow microbes (red) can also exploit fast ones (green). We simulate the evolution of run speed distribution in the presence and absence of sticking, for different channel lengths.}
    \label{fig:model}
\end{figure}

We study an evolving system of actively swimming bacteria in a two dimensional chemostat where a chemoattractant is held at a concentration of $0$ at the left end and increases linearly with constant slope $m$ to reach a maximum value at the right end. The chemostat is a rectangular domain, with Neumann (reflecting) boundary conditions at the walls. In the beginning of each race, bacteria are initialized at the low concentration end of the chemostat, at a fixed distance away from the zero point, and perform run-and-tumble chemotaxis towards the high concentration end, while subject to inter-microbial hydrodynamic interactions, sticking and drag (Fig. \ref{fig:model}). Below we describe how we account for these factors that govern the motion of microbes, and how these factors select for stickiness and run speed over repeated races.

The chemotaxis of microbes is implemented as follows. Every microbe stores a history of chemical concentration of the last 20 time steps, which is then convolved with a response function $K(t)$ to determine the tumble probability $\tilde{\omega}(t)$ at a given time $t$,
\[
\tilde{\omega}(t)= \omega \left[ 1-\int_{0}^{t}K(t-s)c(s)ds \right] .
\]
This equation, together with the chemotaxis response function, is taken from the rigorous experiments of \cite{celani2010bacterial}. Here $c(t)$ is the surrounding nutrient concentration of the bacteria at time $t$, and the chemotactic response $K(t)$ is
\[
    K(t) = \kappa \lambda e^{-\lambda t} \left[ \beta_1 (\lambda t) + \beta_2 (\lambda t)^2 \right],
\]
The response kernel takes the shape of a positive lobe followed by a negative lobe so that a swimming microbe essentially takes a difference between more recent and less recent concentrations to determine whether it is swimming towards a better place. 
 
{\bf Hydrodynamic and adhesive interactions between microbes.} We assign each microbe a hydrodynamic radius $R_h$ and a sticking radius $R_s$. If two bacteria are within a distance $R_h$ from each other, we add the flow velocity generated by each bacteria to the other's motion. Each bacteria generates a dipole flow around itself given by \cite{lauga2009hydrodynamics} 
\begin{equation}
    \label{eq:dipole}
    \mathbf{u}(\mathbf{r}) = \frac{p}{8 \pi \eta r^3} \left[ 3 \cos^2 \theta - 1 \right] \mathbf{r},
\end{equation}
where $p$ gives the strength of the dipole flow and $\eta$ corresponds to the viscosity of the surrounding fluid. The flow strength $p \sim v$ scales linearly with the run speed of the microbe. We therefore simulate the dipole flow generated by a bacteria moving at speed $v_i$ as $(\tilde{p} v_i /r^2) [3 \cos^2 \theta - 1] \mathbf{r}$, where $\tilde{p}$ gives the rescaled strength of the dipole flow for all microbes. Here we only consider pusher microbes, and therefore take $\tilde{p}>0$. Equation (\ref{eq:dipole}) will lead to microbes swimming in the same direction to attract, whereas those swimming in opposite directions will repel. 

In addition to attracting and repelling, microbes also exert a torque on each other from a distance. This torque scales as $1/r^3$, and generally results in microbes aligning parallel with each other if they are pushing swimmers and aligning anti-parallel if they are pulling swimmers. Since we only consider pushers here, and since $1/r^3$ falls of rapidly, we take into account torques pragmatically, by simply aligning the velocity of two microbes that come within distance $R_s$ of each other.

{\bf Aggregation.} Bacteria carry a stickiness trait $s_i$, which takes values of either $0$ (non-sticky) or $1$ (sticky). If two microbes come within a distance $R_s$ they stick together if both of them are sticky; they do not stick if both of them are non-sticky; and they do not stick if one is sticky and the other is non-sticky. This assumption is a special case of that explored in \cite{dufrene2015sticky} (in our case, stickiness is a binary trait rather than a continuous one).
 
We have also explored what happens if the latter of these assumptions is modified (i.e. when a sticky microbe collides with a non-sticky one, they stick) and found that this does not make a qualitative difference in any of our results (cf. Appendix \ref{app:alt_stick}).

Once two microbes stick, we assume that they swim together until one of them tumbles. Stuck pairs move at a modified speed determined from the motile forces exerted by each microbe and the drag forces experienced by the pair, which we discuss below.

{\bf Drag force.} For low Reynolds number, which is the typical environment for bacteria \cite{lauga2009hydrodynamics}, the drag force on a sphere is given by the Stokes law, $D = 6 \pi \eta R v$, where $\eta$ is the fluid viscosity, $R$ the radius of the sphere, and $v$ the velocity of the sphere relative to the fluid. The drag experienced by a pair of spheres is less than twice the drag force experienced by a single sphere, since the stuck pair has less total contact area with the liquid \cite{lee2019hydrodynamic}. We model the effect of a reduced drag by the factor $\gamma$ and take $\gamma$ from earlier theoretical and empirical studies. Since the microbes are still exerting the same force when stuck, the pair will accelerate to reach a new terminal velocity given by 
 \[\mathbf{v}_{\text{pair}} = \frac{\mathbf{v}_1 + \mathbf{v}_2}{ 2 \gamma} .\]
 
In general $\gamma$ will depend on the shape and orientation of the microbes, and can be viewed as a general `cooperative factor' for sticking. Most of our figures are generated  setting $\gamma = 0.7331$, taken from \cite{filippov2000drag} with the assumption that the two microbes are spheres stuck along an axis perpendicular to the direction of motion. However, we also briefly explore the effects of varying this parameter.

{\bf Evolutionary dynamics.} Bacteria reproduce at a rate determined by their local nutrient concentration, cost of moving, and cost of sticking. Specifically, the reproduction rate of bacteria $i$ at position $x_i$ is given by
\[f_i = a \frac{c(\mathbf{x_i})}{c(\mathbf{x_i}) + d} - b v - c s_i\]
where $a$ is the benefit received by the nutrient $c(\mathbf{x})$, $b$ the cost of moving, and $c$ is the cost of being sticky. If $f_i \Delta t$ is negative, the bacteria dies with probability $f_i \Delta t$, if $f_i \Delta t$ is positive, bacteria will reproduce with a probability given by $f_i \Delta t$.

The first term in fitness is a Monod growth function that is empirically verified and commonly used in ecological modelling \cite{monod1949growth, gibson2018distribution}. If the chemoattractant (e.g. nutrition) concentration is much above $d$ (which we set to 1 throughout) the microbe receives diminishing returns. The second term in fitness assumes that the energetic cost of swimming is proportional to velocity. Since the microbe is working against fluid drag and since fluid drag is proportional to velocity in this physical regime, this assumption is reasonable. The last term is essentially an added constant $c$ for sticky microbes and $0$ for non-sticky ones. Here $c$ would be the amount of slowdown in growth rate due to assembling sticky surface glycoproteins or pili, or secreting extracellular polymer substances.

To eliminate discrete-time artefacts, fitness constants and time steps are chosen such that $|f_i \Delta t| \ll 1$. When a cell divides, a new microbe is placed a distance $R_s$ in a random direction away, with a random swim direction, and zero history of past chemical concentrations. The run speed $v$ and stickiness $s$ is inherited. However a random mutations can alter either. Mutations occur at a rate $\mu_v$ for velocities and $\mu_s$ for stickiness. A mutation updates the current velocity by an amount $\delta$ picked from a normal distribution with mean 0 and variance $\sigma_v$. A mutation on stickiness toggles $s_i$ from 0 to 1 or vice versa. 

We simulate multiple races. After a pre-specified race duration $\tau$, a fixed number $n_0$ of randomly chosen bacteria are reset to their original position in the chemostat, as would be during the dilution step of an evolution experiment. Bacteria are placed at the location corresponding to chemical concentration $c_0$ along the horizontal x-axis and uniformly along the vertical y-axis. The repeated races take place up until a total run time $T$. 

A summary of physical parameters is given in Table \ref{tab:model_parameters}. Parameter values were chosen to fit typical values observed for run lengths \cite{berg1972chemotaxis}, bacteria sizes \cite{young2006selective}, and growth kinetics \cite{gibson2018distribution} for bacteria populations. 

\begin{table}
\centering
\begin{tabular}{l l l}
Parameter & Definition & Value \\
\hline
$a$             & Sugar benefit constant    & $40 \times 10^{-3}$ \\
$b$             & Cost of moving            & $0.3 \times 10^{-3}$\\
$c$             & Cost of being sticky      & (0 to 0.36) $\times 10^{-3}$ \\
$m$             & Slope of nutrient concentration & (4.5 to 9) $\times 10^{-4}$\\
$c_0$           & Minimum nutrient concentration & 0.1 \\
$\omega$        & Tumble rate               & 0.1 \\
$\lambda$       & Response time scale       & 0.5 \\
$\beta_1$       & Response shape parameter  & 2 \\
$\beta_2$       & Response shape parameter  & -1 \\
$\kappa$        & Response scaling factor   & 50 \\
$R_s$           & Microbe sticking radius   & 20 \\
$R_h$           & Hydrodynamic radius       & 50 \\
$\gamma$        & Hydrodynamic drag factor  & 0.5 to 1.0 \\
$\tilde{p}$     & Hydrodynamic dipole factor & 50\\
$s$             & Microbe stickiness        & 0, 1.0 \\
$\mu_v$         & Velocity mutation rate    & $1 \times 10^{-4}$\\
$\sigma_v$      & Velocity mutation strength & 1.0 \\
$\mu_s$         & Stickiness mutation rate  & $1 \times 10^{-5}$ \\
$n_0$           & Number microbes reset     & 1000 \\
$\tau$          & Race duration             & 10 to 300 \\
$T$             & Total evolutionary duration            & 50000 \\
$\Delta t$      & Time step                 & 1 \\
$H$             & Domain height             & $2 \times 10^3$ to $2 \times 10^7$ \\
$W$             & Domain width              & $2 \times 10^5$ 
\end{tabular}
\caption{A summary of the model parameters and the default values used in simulations.}
\label{tab:model_parameters}
\end{table}


Before we move on to describing our results, we should warn that in all of our simulations, we consider only pairwise interactions between microbes. This means that our results are valid only when the microbial swarm is moderately sparse. More specifically, our model will hold true if the number of instances where 3 or more sticky microbes happens to be within $R_h$ (and thus $R_s$) is negligible compared to the number of instances where a radius of $R_h$ contains one or two sticky microbes.

\section*{Results}

\subsection*{Optimal velocity in the absence of hitchhiking}
\begin{figure}
    \centering
    \includegraphics[width=0.43\textwidth]{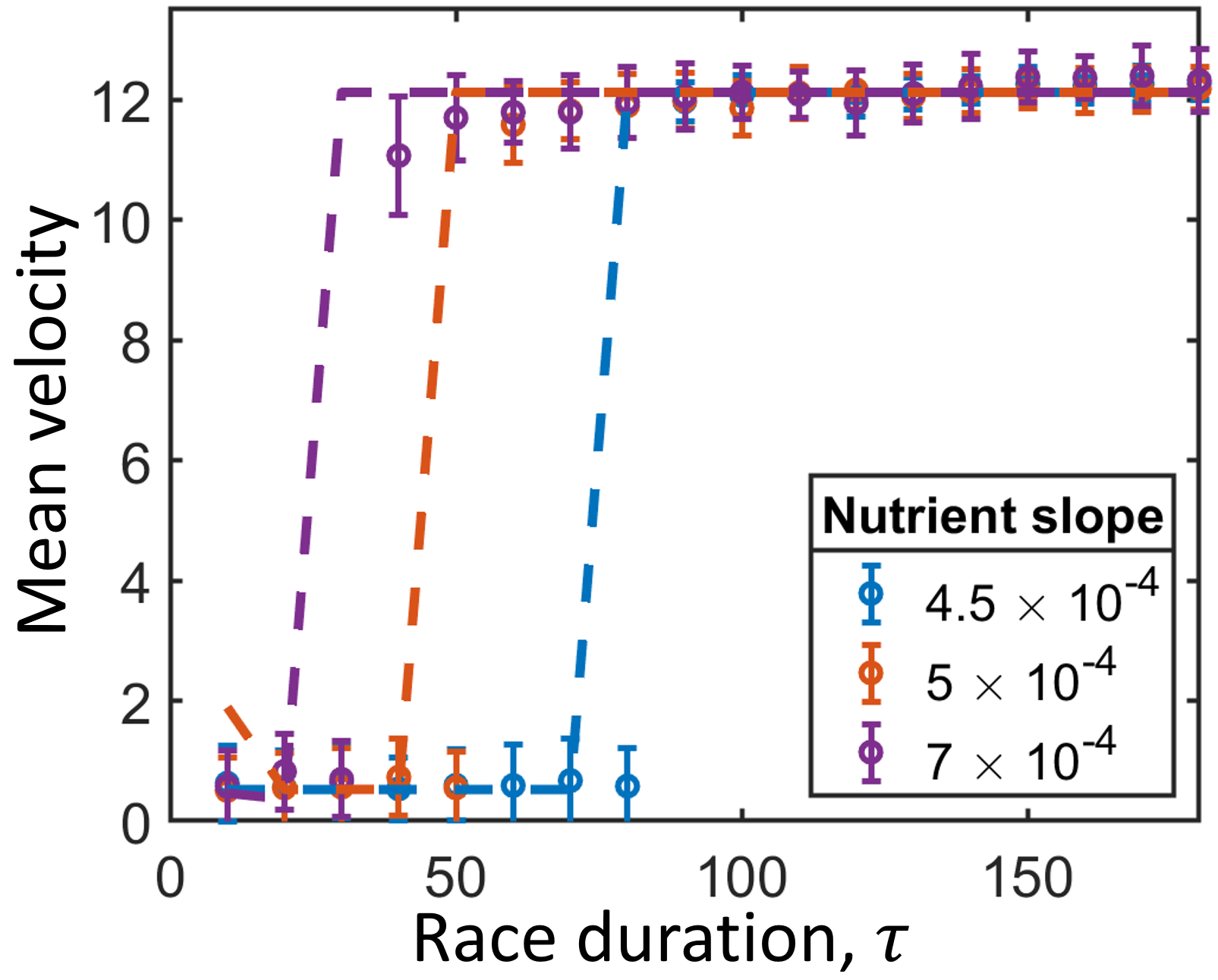}
    \caption{{\bf Optimal velocity versus race duration without hitchhiking.} Mean optimal velocity versus race duration for different nutrient slopes. In shorter races, faster moving microbes pay a larger cost and do not gain as much of an advantage from moving. In longer races, the optimal mean velocity increases to a saturating value given by the maximum viable velocity which can be sustained at the minimum nutrient concentration $c_0$ (equation (\ref{eq:vmax})). As we vary the nutrient slope, we effectively rescale space. For larger slopes, the benefits of moving are realized at shorter races. We can determine the optimal mean velocity analytically (equation (\ref{eq:vmean})). Points are from simulation data ran for a total duration of $T=50000$ and dashed lines from equation (\ref{eq:vmean}). Numerical results are independent of the initial velocity distribution. The mean speeds are therefore evolutionarily stable strategies. The chemostat width is $H = 2000$, and the rest of the parameters are as given in Table \ref{tab:model_parameters}. }
    \label{fig:velocity_runtime}
\end{figure}

We first determine the evolutionarily optimal swimming speeds when there is no cell-cell sticking (and no stickiness cost). We ran simulations varying the race duration $\tau$ and nutrient slope $m$. Overall we observe a uni-modal distribution with the mean velocity increasing to a maximum optimal value for longer races (Fig. \ref{fig:velocity_runtime}). This optimal value is independent of the initial velocity distributions and is evolutionarily stable. 

For very long races, one might guess that faster is always better, since those that reach the high end of the chemogradient early on will have the most offspring. However this is not the case. Microbes that swim too fast cannot recover the energy they expend while they are at the low end of the gradient, thus, the optimal velocity for large race durations is determined by the maximum viable run speed at the beginning of each race. That is,
\begin{equation}
    v_{\text{max}} = \frac{a}{b} \frac{c_0}{c_0 + d}.
    \label{eq:vmax}
\end{equation}
Mutations may allow larger velocities to emerge once slow microbes reach higher nutrition values, but these faster swimming microbes will die out in the beginning of the next race.

Therefore, in short races, microbes do better by not swimming. Beyond a critical race duration $\tau>\tau_c$, it becomes best for microbes to swim at their maximum viable run speed given by equation (\ref{eq:vmax}). We also obtain this critical transition time $\tau_c$ analytically as given by dashed lines in Fig. \ref{fig:velocity_runtime}, and derived in Appendix \ref{app:derivations} and get good agreement with simulations.

\subsection*{Effects of hitchhiking for fixed stickiness}

We now study the effect of sticking on the optimal swimming speed of bacteria. We first investigate the case where sticking has no fitness cost, and where everyone has the same stickiness.

\begin{figure*}
    \centering
    \includegraphics[width=0.94\textwidth]{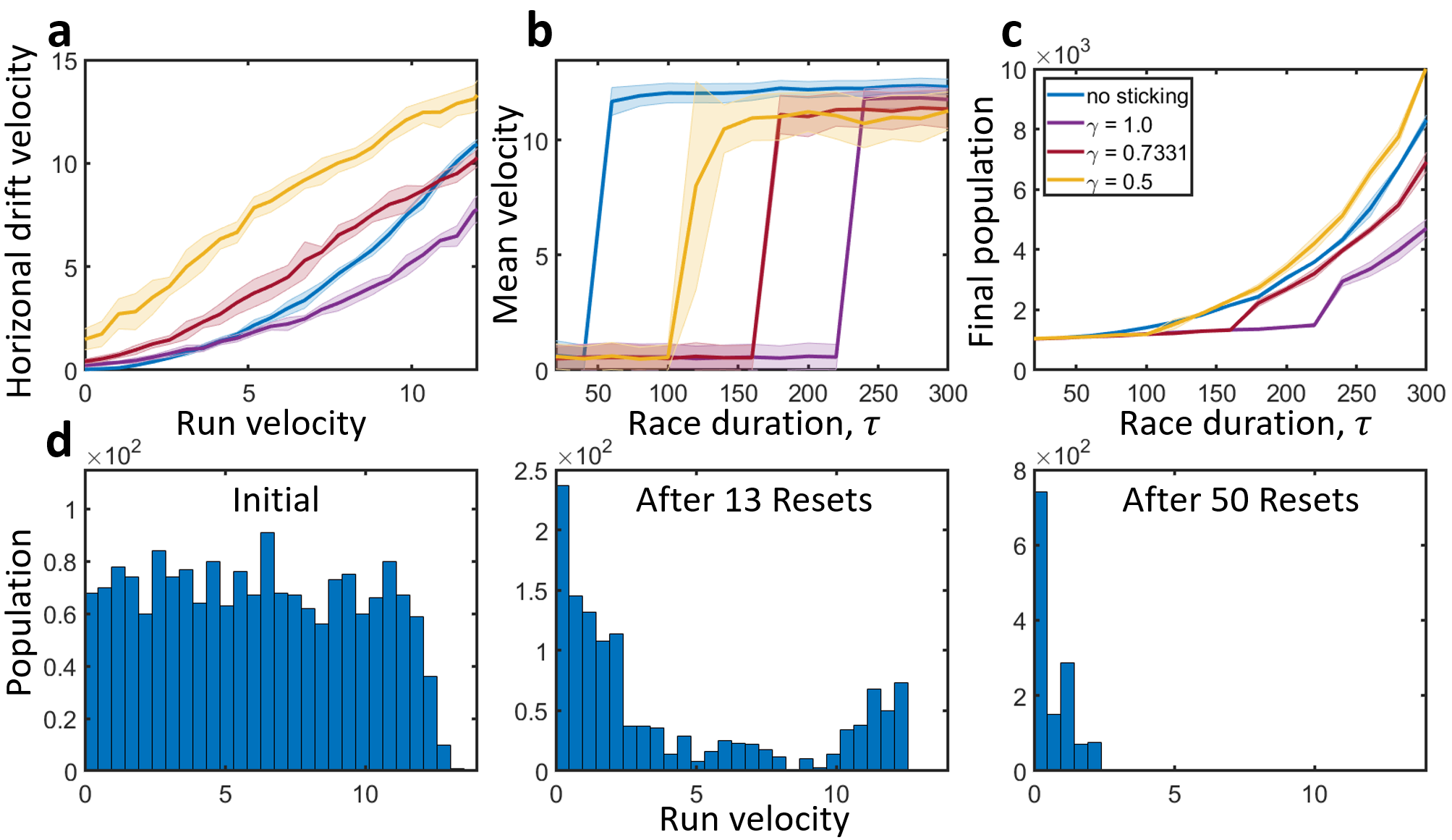}
    \caption{{\bf Effects of sticking.} {\bf a,} Effect of stickiness on drift velocity for various drag factors $\gamma$. The solid blue line gives the drift velocity in the $x$ direction versus run speed in the absence of sticking. The purple line corresponds to sticking with $\gamma=1$. Here stickiness gives a boost to the slowest microbes allowing them to move faster than would be possible on their own, and slows down fast moving microbes. As we decrease the factor $\gamma$, microbes with a larger range of velocities benefit from sticking. For $\gamma=0.5$, we see that everyone moves faster than the non-sticking case. The drift velocity was found by taking the final displacement of a population after a time $T = 300$, starting with a uniform initial distribution of run velocities. {\bf b,} Optimal mean velocity versus race duration for non-sticking (blue) and sticking populations. For short races, stickiness does not affect the optimal mean velocity. At intermediate race durations, stickiness allows slow microbes to dominate where they otherwise would not. For long races, fast microbes can cooperate through sticking and lower their mean velocity compared to the case without sticking. As we lower the drag factor $\gamma$, sticking becomes advantageous to fast microbes at shorter race durations. {\bf c,} Final population after $T = 50000$ for non-sticking (blue) and sticking populations for different drag factors. For short races, stickiness does not effect the final population. For intermediate-length races, stickiness allows slow microbes to reach larger nutrient concentrations and dominate the population. This leads to a tragedy of the commons and the final population is lower than without stickiness. For long races, fast microbes cooperate through sticking and the final population increases. As we lower $\gamma$, the region where the tragedy of commons shrinks and the cooperative region where a sticky fast population outperforms a non-sticky one comes at an earlier race duration. {\bf d,} Evolution of velocity distribution for intermediate-length race with $\tau = 160$ and drag constant $\gamma = 0.7331$. An initially uniform velocity distribution becomes transiently bimodal as slow microbes exploit fast ones to move to larger nutrient regions. Finally, the slow microbes dominate, leading to a tragedy of the commons where there are no longer fast microbes for slow ones to exploit.}
    \label{fig:stick_effect}
\end{figure*}

We find, starting from an initially uniform velocity distribution, sticking mostly benefits slower moving microbes, giving them the largest velocity boost, and the fastest microbes are harmed by being slowed down from sticking to slower microbes (Fig. \ref{fig:stick_effect}a). As we lower the drag factor $\gamma$, a larger proportion of run velocities is benefited by sticking and at $\gamma=0.5$ we see that everyone moves faster through sticking than without. This effect then benefits the slower microbes best at intermediate-length races, $\tau$ (Fig. \ref{fig:stick_effect}b). For shorter races, slow microbes already do the best. For long races, slow microbes gain an initial advantage, but eventually fall behind. Fast microbes on their own move faster than pairs of slow and fast microbes, and thus still dominate in long races. For intermediate races however, slow microbes are able to make it to regions of larger nutrient concentration without expending as much energy as fast swimmers. Over the course of many repeated races, the population distribution transiently becomes bimodal and slow microbes benefit from hitchhiking on fast ones. Eventually the population becomes dominated by slow microbes (Fig. \ref{fig:stick_effect}d). This is a typical ``tragedy of the commons'' scenario, where the cheating strategy takes over and fast microbes no longer exist to help disperse slow microbes. To see this clearly, we plot the population for sticking and non-sticking populations versus race duration in Fig. \ref{fig:stick_effect}c. For intermediate races, the population decreases from sticking, since slow microbes cause a tragedy of the commons. In long races, fast microbes are able to cooperate with each other via sticking and the overall population increases compared to a non-sticking population. As the drag factor $\gamma$ decreases, this region of tragedy of the commons shrinks and the region where fast microbes benefit by cooperating and sticking comes at an earlier race duration.

\begin{figure}
    \centering
    \includegraphics[width=0.49\textwidth]{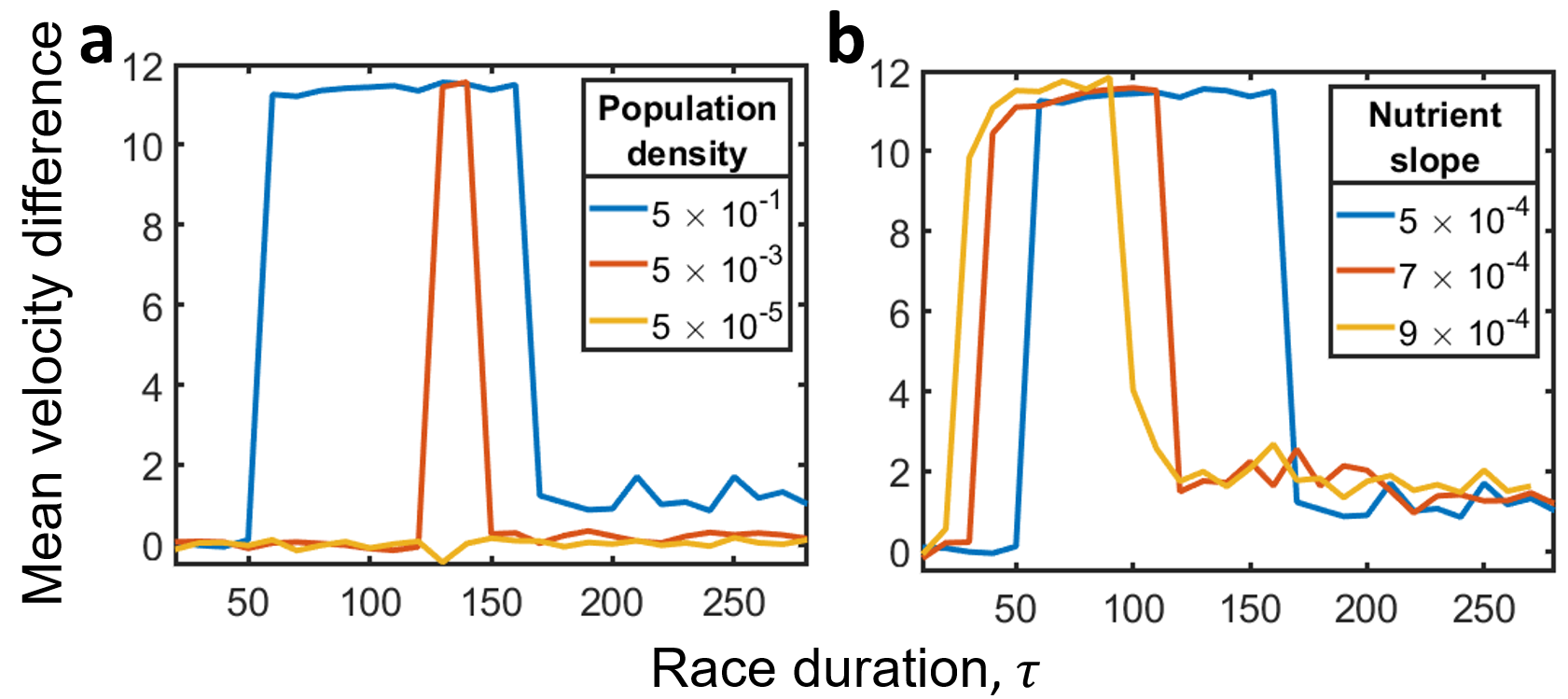}
    \caption{{\bf Coupled effects of sticking, density, and nutrient slope.} {\bf a,} Mean velocity difference versus race duration for various population densities. The mean velocity difference is given as the optimal mean velocity without stickiness minus the optimal mean velocity with stickiness set to one. The velocity difference peaks at intermediate race durations. For long races there is a small velocity difference from fast microbes cooperating to lower their mean velocity to a new optimum. As the population density decreases, the effect of sticking diminishes, and the race duration region where stickiness benefits slow microbes shrinks. {\bf b,} As we vary the slope, the position of the peak shifts. A larger slope shifts the position of the peak to shorter race durations. The width of race durations where sticking makes a difference also shrinks with larger slope. The effect of varying slope can essentially be understood from a rescaling of space.}
    \label{fig:coupled_stick}
\end{figure}

We next study the effects of stickiness for varying microbial density and nutrient gradients. To tune the microbial density, we varied the height $H$ of the simulation domain, keeping the number of bacteria at the beginning of each reset $n_0$ constant. We plot the mean velocity difference -- the mean velocity when there is no sticking minus the mean velocity when stickiness is one -- versus race duration, for varying density and nutrient slope in Fig. \ref{fig:coupled_stick}. The plots show a maximal difference at an intermediate race duration as discussed above.

When we vary the microbial density, we find the peak becomes less wide for sparser populations. For denser populations, sticking events are more frequent and the effect of stickiness is more pronounced. 

When we vary the slope we see the peaks shift to different race durations. For higher nutrient slopes, the peaks occur at shorter race durations. This is due to the shift in the transition race duration seen in the case without sticking (Fig. \ref{fig:velocity_runtime}). This can also be seen as a rescaling of space. A larger slope brings the high nutrient concentration region closer, and so at larger nutrient slopes the benefit of swimming is realized at shorter races. The advantage from sticking is therefore also realized at shorter races, and the peaks shift towards lower race durations at larger slopes. 

Thus, we see that the optimal conditions for employing a sticking strategy vary with population density and nutrient slope. The slow and sticky cheaters are better off always at intermediate-length races, which can be interpreted as sparse nutrient concentrations and/or large consumption and decay rates of nutrients. Next we show how sticking strategies may evolve naturally for microbes for varying costs and hydrodynamic drag factors associated with sticking. 

\subsection*{Coevolution of run speed and stickiness}

\begin{figure}
    \centering
    \includegraphics[width=0.49\textwidth]{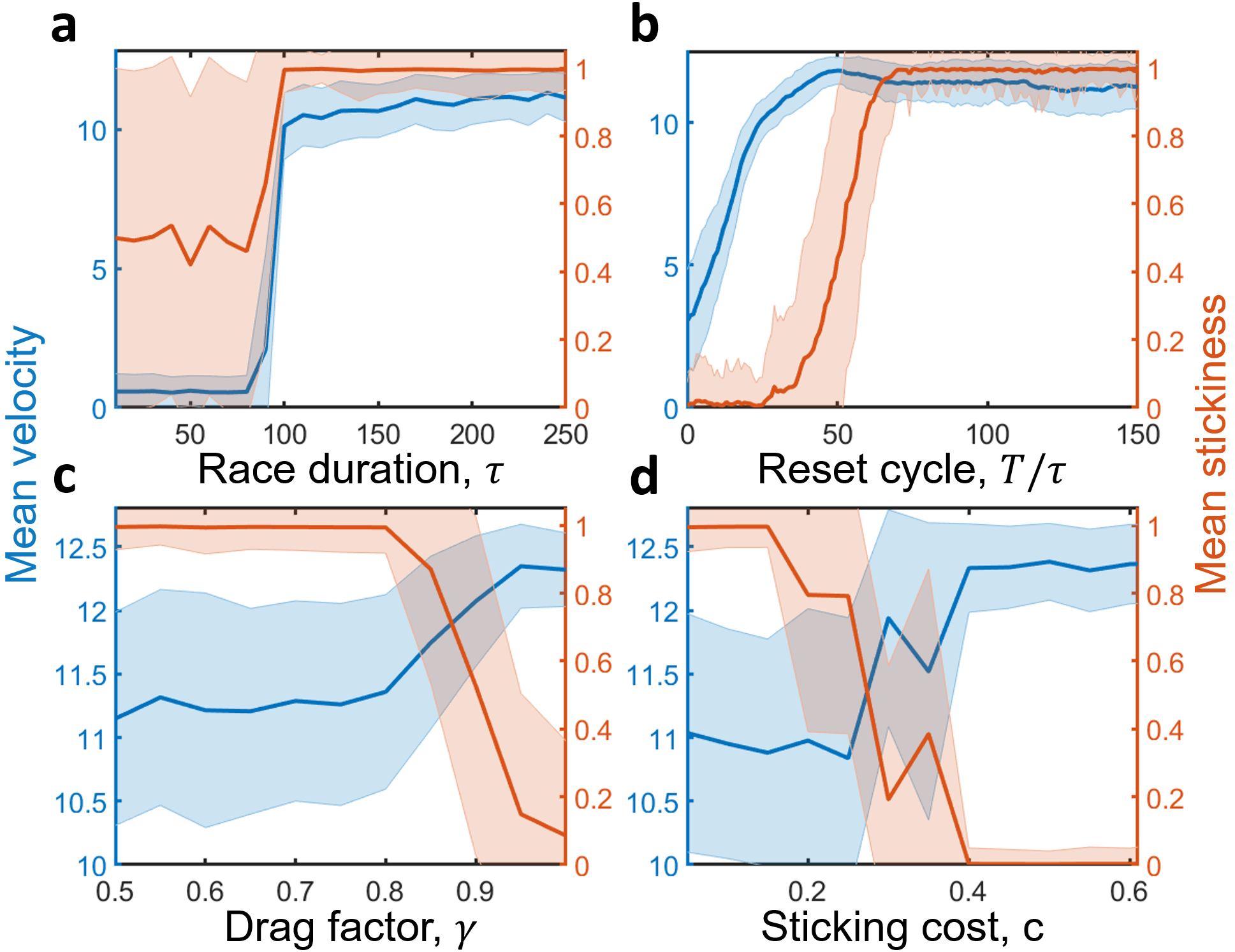}
    \caption{{\bf Evolution of stickiness.} {\bf a,} For short races, microbes evolve to swim at slower velocities. Here, stickiness is neutral and evolves to be around $\langle s \rangle = 0.5$ on average due to genetic drift. For long races, microbes are narrowly distributed around a faster velocity distribution. Here sticking allows microbes to cooperate and move faster than alone. Stickiness therefore evolves to be around $\langle s \rangle = 1.0$. {\bf b,} Evolution of run speed and stickiness for long race durations ($\tau = 200$). Here, an initially uniform population quickly evolves to have a large mean velocity. It then becomes advantageous to stick. At this point, microbes evolve to become sticky and lower their run speed to out-compete non-sticky, faster microbes. {\bf c,} Mean velocity and stickiness versus drag factor $\gamma$ for $\tau = 200$ and zero cost. For long race durations, the population predominately consists of fast microbes. Since nearby microbes are close to the same velocity, for low drag factors $\gamma$ stickiness offers an advantage to microbes. They therefore evolve to be sticky and can lower their velocity and out-compete faster microbes. The mean velocity therefore decreases slightly. For a larger drag factor, stickiness actually harms microbes since sticking to a randomly moving microbe slows it down on average. Around $\gamma = 0.82$, microbes evolve to not be sticky and move on their own at a larger run speed. {\bf d,} Mean velocity and stickiness versus sticking cost for $\tau = 200$, and $\gamma = 0.6$. As the cost of stickiness increases, there is a trade-off between sticking to boost the drift velocity and moving on one's own without expending resources to stick. Once the cost of sticking is too large, microbes evolve to not stick and swim at a larger velocity instead. }
    \label{fig:stick_mutation}
\end{figure}

Finally, we explore how microbes may adapt their sticking strategies by allowing stickiness to mutate. We determine how sticking strategies may evolve over time and the effects of reduced drag $\gamma$ and sticking cost. One method of sticking together is through the use of secreted extracellular substances. These substances may be costly to produce, but advantageous to slow cheaters or mutually cooperating fast microbes. We therefore add an associated cost $c s_i$ to sticking. The probability for two bacteria to stick $p$ is given by the product of the two bacteria's stickiness constants $p = s_1 s_2 \in \{0,1\}$. We also explore the alternative case where $p = \max (s_1, s_2) \in \{0,1\}$ in Appendix \ref{app:alt_stick} and find no qualitative differences.

We first study stickiness evolution for varying race durations and plot the mean stickiness and mean velocity (Fig. \ref{fig:stick_mutation}a). We find that when there is no cost, slow microbes evolve to an average stickiness close to $0.5$ for short races. This is the case whether we start with an initial population of all stickiness or near zero stickiness. Here, since the population is composed of essentially non-motile microbes, stickiness does not have a significant effect. Hence, the stickiness of microbes evolves primarily due to genetic drift. For long races, the population is composed of faster microbes. Since the velocity distribution is concentrated around fast microbes, sticking helps fast microbes as they stick to other fast microbes and reduce their drag force. We see in Fig. \ref{fig:stick_mutation}b, that microbes evolve to be sticky after they have evolved to have fast velocities. Here, the stickiness of fast microbes evolves to near one and the mean velocity slightly drops to where slower microbes out-compete the very fast ones. Hence, what is seen as a parasitic interaction between slow and fast microbes, becomes a cooperative interaction between fast microbes themselves. Since the population consists of predominately fast microbes, sticking is mutually beneficial in long races. 

The amount by which sticking helps microbes will in general depend on their shape and hydrodynamic properties. In Fig. \ref{fig:stick_mutation}c, we plot the mean velocity and stickiness as a function of the drag factor $\gamma$. For lower values of $\gamma$ sticking allows pairs of microbes to reduce their hydrodynamic drag and increase their drift velocity. Fast microbes therefore evolve to become sticky. At larger values of $\gamma$, sticking no longer becomes beneficial, and in fact begins to slow microbes down as they stick to other microbes moving in random directions. Therefore, around $\gamma = 0.82$ (Fig. \ref{fig:stick_mutation}c), microbes evolve to lose their stickiness. Finally, we study the effects of having a sticking cost, with the drag factor fixed at $\gamma = 0.6$. Even with some cost, sticking offers a larger advantage to microbes. Once the cost becomes too large however, around $c =0.2$, sticking no longer becomes beneficial and microbes evolve to lose stickiness, and increase their mean velocity instead (Fig. \ref{fig:stick_mutation}d).

\begin{tcolorbox}[enhanced] 
\label{box:summary}
{\bf Cheating and tragedy of commons:} At intermediate distances between nutrient patches, slow moving cheaters gain the most benefit by sticking to faster microbes to move to ``greener pastures'' without expending effort on their own. Over many evolutionary runs, parasitic slow microbes out-compete fast ones, leading to a tragedy of the commons where there are no longer fast microbes left to exploit. The final population of microbes is then lower compared to the non-sticking case (Fig. \ref{fig:stick_effect}c). \\
{\bf Hydrodynamic cooperation:} For long races, fast microbes leave slow ones behind. They can then cooperate with each other by sticking and reducing their hydrodynamic drag. Sticking fast microbes then do better compared to non-sticking (Fig. \ref{fig:stick_effect}c). When allowing sticking to be a mutable trait, we see fast microbes naturally evolve to stick at long race durations and sufficiently low drag and cost to sticking (Fig. \ref{fig:stick_mutation}).
\end{tcolorbox}
\vspace{2mm}
\section*{Discussion}
The phenomena of hitchhiking has been observed experimentally, but a theoretical understanding of its evolution and ecological function has been lacking. Here we study a simple model in which slow microbes can stick to faster ones to hitch a ride for free, as well as faster microbes sticking together to mutually benefit from reduced drag. 

In addition to aggregation, we also accounted for hydrodynamic forces in the evolution of microbial motility. Specifically we investigated the effects of self generated flows and reduced drag forces experienced by pairs of microbes. We also accounted for the drag force modification factor $\gamma$ (which will depend on the shape and orientation of microbes) and studied its role in the evolution of different motility strategies.

We ran ``evolutionary experiments'' where microbes actively swim up a nutrient gradient for a predetermined race duration. The race duration can be interpreted as the average distance between nutrient patches or the decay time of transient nutrition concentrations. After this decay time the chemical concentration is reset and microbes swim to the next patch.

Through our first-principle simulations, we find that when nutrients are distributed at short distances, the best strategy is for no one to swim. At intermediate nutrient distributions, slow microbes evolve to hitchhike on faster ones, leading to a tragedy of commons where there are no longer fast microbes left to exploit. When nutrient sources are distributed far apart, fast microbes evolve to adhere to each other to cooperate and reduce their hydrodynamic drag, benefitting the whole population.

Sticking therefore goes from meditating a parasitic interaction, leading to a tragedy of commons, at intermediate nutrient distributions, to an evolutionarily stable mutualistic interaction amongst fast microbes when nutrients are scarcely distributed. We therefore find the ecological nature of hitchhiking will depend on the nutrient landscape and on the hydrodynamic drag forces on microbes, which are related to microbial shape and orientation.

Throughout, we paid close attention to physical realism, however we also made important simplifying assumptions. To simplify our analysis and to capture the relevant phenomena, we implemented evolutionary simulations in a controlled chemostat environment with a linearly increasing chemical profile. We also focused on pair-wise interactions between microbes. Other mechanisms may also contribute to the aggregation of microbes and would be interesting to investigate in this context. For example, turbulent forces can cause accumulation of cells. This effect also depends on the shape of microbes \cite{zhan2014accumulation}. 


\bibliographystyle{unsrt}
\bibliography{bibliography}

\appendix
\renewcommand{\figurename}{Supplementary Figure}
\setcounter{figure}{0}  

\section{Semi-analytical results for optimal velocity in the absence of hitchhiking}
\label{app:derivations}

Here we derive semi-analytical results for determining the critical time $\tau_c$ where swimming becomes advantageous to microbes.



We can describe the run-and-tumble motion of a population at scales larger than the run length and time scales longer than the tumble time, via an effective diffusion-advection equation. Adding in mutations and reproduction terms, we can effectively describe the model with a continuous system of partial differential equations for non-sticky microbial density $n = n(x,y,v)$ as,
\[
    \dot{n} = \left( D \nabla^2 - \varepsilon v \partial_x + \mu_v \sigma_v^2 \partial_v^2 + a \frac{mx \!+\! c_0}{mx \!+\! c_0 \!+\! 1} - b v  \right) n 
\]
where the effective diffusion $D$ and chemotactic efficiency $\varepsilon$ will depend on the response kernel $K(t)$. The effective diffusion is simply given as $D = v^2 / 2 \omega$. Following the procedure given in \cite{celani2010bacterial}, we also obtain an expression for the chemotactic efficiency $\varepsilon$,
\[
    \varepsilon = \frac{m \kappa v}{2 \omega} \left[2 \beta_2 \frac{\lambda^3}{(\lambda + \omega)^3} + \beta_1 \frac{\lambda^2}{(\lambda + \omega)^2} \right] .
    \label{eq:epsilon}
\]

To determine the mean velocity versus race duration theoretically, we first simplify our system by ignoring diffusion and mutations, and assume everyone moves at a velocity $\tilde{\varepsilon} v$. Here, due to additional hydrodynamic interactions as well as the effects of diffusion and reproduction, we have $\tilde{\varepsilon} > \varepsilon$ as given in equation (\ref{eq:epsilon}), since alignment generally helps orient velocities towards nutrients and the growth rate of microbes that diffuse ahead of the mean is larger than those that fall behind. This value is not straightforward to obtain theoretically because of the saturated growth. We therefore measure this quantity from simulations.

We then obtain an ordinary differential equation describing the growth of the population $n(v,t)$, 
\[
    \dot{n} = n \left[ a \frac{m \tilde{\varepsilon} v t + c_0}{m \tilde{\varepsilon} v t + c_0 + 1} - b v \right].
\]

We can solve this analytically to get,
\[
    n_1 = n_0(v) e^{t (a-b v)} \left[ \frac{c_0+1}{c_0 + 1 + \tilde{\varepsilon} m t v} \right]^{a/ \tilde{\varepsilon} m v} .
\]

Where $n_0 (v)$ is the initial velocity distribution. We can describe the result of restarting the run $N = T/\tau$ times, by taking the distribution at the end of a race as the initial distribution and repeating the process, times a normalization factor. Therefore, after $N$ iterations, the distribution asyptotically approaches,
\begin{equation}
    n_N(v, \tau) \sim e^{N \tau (a-b v)} \left[ \frac{c_0+1}{c_0 + 1 + \tilde{\varepsilon} m \tau v} \right]^{a N / \tilde{\varepsilon} m v} .
\label{eq:popevo}
\end{equation}

We can then get the mean velocity after $N$ resets and race duration $\tau$ by taking the average,
\begin{equation}
    \langle v \rangle_N (\tau) = \frac{\int_0^{v_{\text{max}}} v n_N(v,\tau) \ \mathrm{d} v}{\int_0^{v_{\text{max}}} n_N(v,\tau) \ \mathrm{d} v} .
\label{eq:vmean}
\end{equation}
We compare this to simulation results in Fig. \ref{fig:velocity_runtime} and get good agreement.

We note in equation (\ref{eq:popevo}), as race duration $\tau$ goes to infinity, the optimal velocity goes to zero, since any small positive velocity will reach high enough saturating goods and out-compete faster microbes. However, in a more natural setting, microbes will consume the resources and slow ones may not actually make it to the resource in time. For shorter races then, there is an advantage to swimming, and the optimal run speed behaves as in Fig. \ref{fig:velocity_runtime}.

\section{Significance of sticking assumptions}
\label{app:alt_stick}
Here we determine the significance of the sticking assumptions made in the paper. Specifically, we explore what happens if we modify our assumption that a sticky and non-sticky microbe do not stick and instead have them stick. We find this modification does not make a qualitative difference in any of our results. 

\begin{figure}
    \centering
    \includegraphics[width=0.48\textwidth]{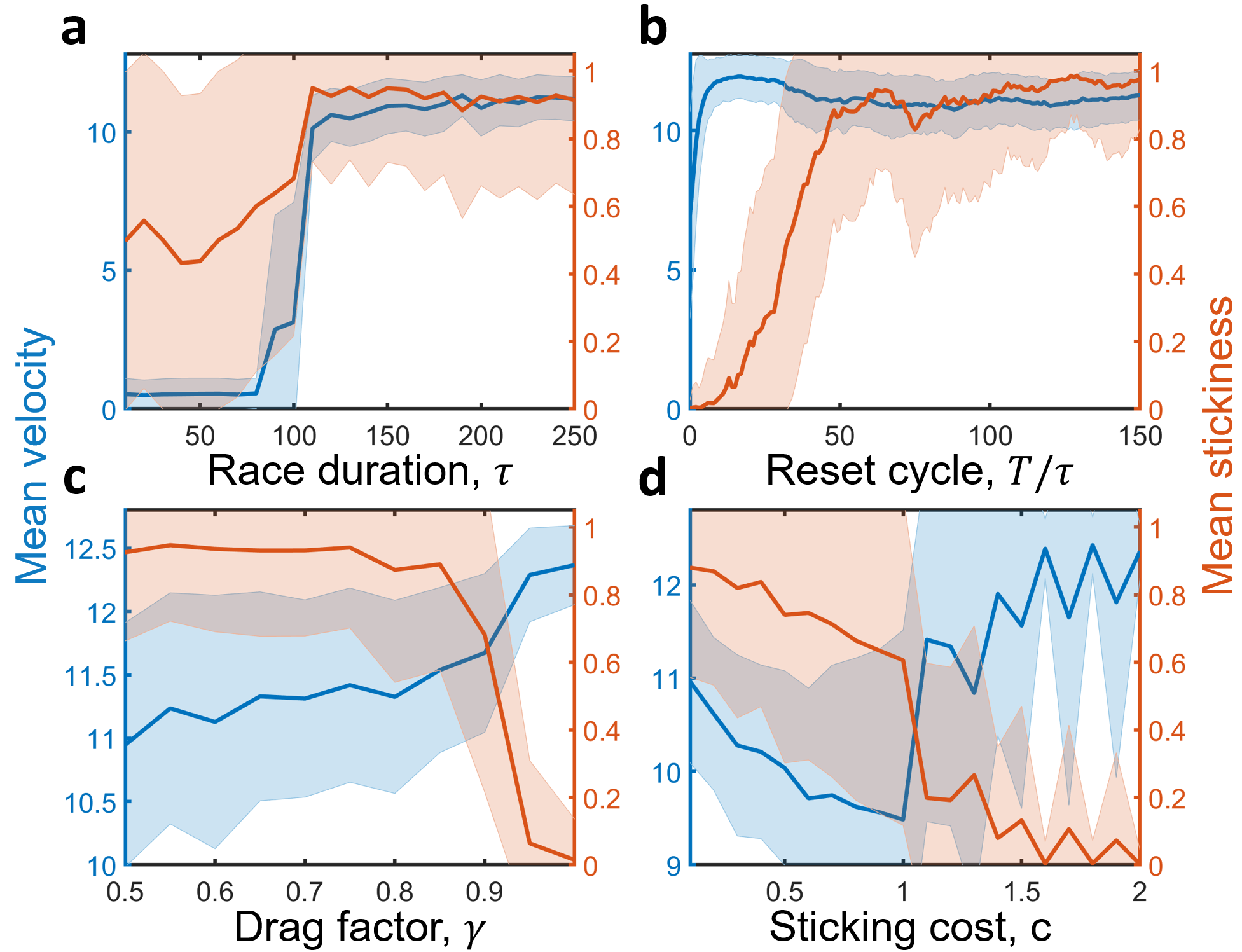}
    \caption{{\bf Evolution of stickiness with modified sticking scheme.} Here we reproduce figure \ref{fig:stick_mutation} in the text with the modified assumption that a sticky and non-sticky microbe stick together when coming into contact. For mean velocity and stickiness versus race duration {\bf a,} over reset cycles {\bf b,} and versus drag factor $\gamma$ {\bf c,} we see no significant quantitative difference. For the mean velocity and stickiness versus sticking cost {\bf d,} we see a quantitative difference but observe the same qualitative behavior. Here since some non-sticky microbes can still hitchhike when coming into contact with sticky ones, the critical cost where the population evolves to loose stickiness is now at a larger value. The transition from sticking to non-sticking is also more gradual compared to the case where sticky and non-sticky microbes do not stick.}
    \label{fig:app_stick_evo}
\end{figure}

For results where sticking is not subject to mutation, the assumption makes no difference at all since all microbes are taken to be either fully sticking or non-sticking. Here the case of interest where a non-sticky microbe encounters a sticking one does not occur.

In the case where we do allow stickiness to mutate, we find no change in our results when varying race duration and drag factor (Fig. \ref{fig:app_stick_evo}a-c). We do see a quantitative change when varying sticking cost (Fig. \ref{fig:app_stick_evo}d), but observe the same qualitative behavior. Here, as we increase sticking cost, a fraction of the population evolves to not be sticky but can still hitchhike due to other sticking microbes. As the cost increases, a larger fraction of the population evolves to be non-sticky until a critical cost where the cost of sticking outweighs the benefit and microbes evolve to be non-sticky and swim alone at a faster speed. Compared to Fig. \ref{fig:stick_mutation}d, we see the critical cost where it is no longer advantageous to stick is now at a larger value and the transition from non-sticking to sticking is more gradual.

\end{document}